\documentclass[conference]{IEEEtran}
\IEEEoverridecommandlockouts
\usepackage[top=0.75in, bottom=1.0in, left=0.625in, right=0.625in]{geometry} % ICC submission edas

\usepackage[utf8]{inputenc}
\usepackage{cite}
\usepackage{amsmath,amssymb,amsfonts}
\usepackage{algorithm}
\usepackage{algorithmic}

\usepackage{textcomp}
\usepackage{xcolor}
\usepackage{siunitx}
\usepackage{booktabs}
\usepackage{multirow}
\usepackage{colortbl}
\usepackage{tikz}
\usepackage{enumitem}
\usepackage{graphicx}
\usepackage{caption}
\usepackage{subcaption}

\usepackage{times}

\usepackage{fancyhdr}
\fancypagestyle{firststyle}{\fancyhf{}
		\fancyhead[L]{M. Ying, G. Qian, X. Wang, P. Ma, D. Shakya, and T. S. Rappaport,``HoRAMA: Holistic Reconstruction with Automated Material Assignment for Ray Tracing using NYURay", to appear in \textit{IEEE International Conference on Communications (ICC), }Glasgow, UK, Jun. 2026, pp. 1--6.}
	}
\def\BibTeX{{\rm B\kern-.05em{\sc i\kern-.025em b}\kern-.08em
    T\kern-.1667em\lower.7ex\hbox{E}\kern-.125emX}}

\begin{document}
\bstctlcite{BSTcontrol}
\title{HoRAMA: Holistic Reconstruction with Automated Material Assignment for Ray Tracing using NYURay\thanks{This work was supported by the NYU WIRELESS Industrial Affiliates Program, NYU Tandon ECE PhD Fellowship, and NSF Grant No. 2234123. The authors thank Prof. David Fouhey and Mr. Linyi Jin for their valuable discussions and suggestions.}}

    \author{\IEEEauthorblockN{Mingjun Ying\IEEEauthorrefmark{1},  Guanyue Qian,  Xinquan Wang, Peijie Ma, Dipankar Shakya, and Theodore S. Rappaport\IEEEauthorrefmark{2}}
    \IEEEauthorblockA{NYU WIRELESS, New York University, Brooklyn, NY, USA}
    \IEEEauthorblockA{\{yingmingjun\IEEEauthorrefmark{1}, tsr\IEEEauthorrefmark{2}\}@nyu.edu,}
}

\maketitle

\thispagestyle{firststyle}

\begin{abstract}
    Next-generation wireless networks at upper mid-band and millimeter-wave frequencies require accurate site-specific deterministic channel propagation prediction. Wireless ray tracing (RT) provides site-specific predictions but demands high-fidelity three-dimensional (3D) environment models with material properties. Manual 3D model reconstruction achieves high accuracy but requires weeks of expert effort, creating scalability bottlenecks for large environment reconstruction. Traditional vision-based 3D reconstruction methods lack RT compatibility due to geometrically defective meshes and missing material properties. This paper presents Holistic Reconstruction with Automated Material Assignment (HoRAMA) for wireless propagation prediction using NYURay. HoRAMA generates RT-compatible 3D models from RGB video readily captured using a smartphone or low-cost portable camera, by integrating MASt3R-SLAM dense point cloud generation with vision language model-assisted material assignment. The HoRAMA 3D reconstruction method is verified by comparing NYURay RT predictions, using both manually created and HoRAMA-generated 3D models, against field measurements at 6.75 GHz and 16.95 GHz across 12 TX-RX locations in a 700 square meter factory. HoRAMA ray tracing predictions achieve a 2.28 dB RMSE for matched multipath component (MPC) power predictions, comparable to the manually created 3D model baseline (2.18 dB), while reducing 3D reconstruction time from two months to 16 hours. HoRAMA enables scalable wireless digital twin creation for RT network planning, infrastructure deployment, and beam management in 5G/6G systems, as well as eventual real-time implementation at the edge~\cite{wang2005parametric}.
\end{abstract}


\section{Introduction}
\label{sec:introduction}

Rapid proliferation of wireless devices and exponential growth in data traffic drive next-generation networks toward upper mid-band (above 6 GHz) and millimeter-wave (mmWave) frequencies to achieve higher data rates and lower latency~\cite{rappaport2013millimeter,Shakya2024ojcoms}. Accurate propagation prediction becomes critical for network planning and infrastructure deployment as well as research and development for futuristic applications such as integrated sensing and communication (ISAC) and position location~\cite{kanhere2018position,kanhere2019map,  bazzi2025isac}. Properly modeling electromagnetic (EM) interactions with building materials and objects significantly impacts coverage and performance and have posed the most time-consuming bottleneck to reliable RT simulations~\cite{ying2025site}. Traditional statistical channel models, such as 3GPP models for urban microcell (UMi) and indoor hotspot (InH) scenarios, provide general propagation characteristics but fail to capture the spatial, temporal, and angular resolution required for site-specific prediction above 6 GHz.

Recent measurement data standards~\cite{Ted2025icc, shakya2025milcom} enable ML tools to train on pooled measurements that incorporate both statistical and site-specific features across diverse scenarios.

Although ray tracing (RT) has been explored for over three decades, \cite{seidel1994tvt,rappaport2005BDM,rappaport2004RTsurface}, the advent of artificial intelligence (AI) and extremely capable graphics processing unit (GPU) makes wireless RT viable as a deterministic approach for site-specific channel prediction, enabling accurate modeling of line-of-sight (LOS) and non-line-of-sight (NLOS) propagation through reflection, penetration, diffraction, and scattering mechanisms~\cite{mckown1991inet,schaubach1992microcellular,sun2014icc,nie2013indoor,wang2005parametric}. RT has gained renewed attention for applications including indoor localization~\cite{kanhere2018position}, digital twin modeling~\cite{Lei2025DTWIN}, and 5G/6G network planning~\cite{hoydis2023sionna}. RT platforms include Remcom Wireless InSite~\cite{RemcomWirelessInSite}, MATLAB Ray Tracing, and NVIDIA Sionna~\cite{hoydis2023sionna}, while implementations such as NYURay demonstrate millimeter-wave channel prediction capabilities through measurement-driven EM material calibration across multiple frequency bands and deployment scenarios~\cite{xing2021millimeter,kanhere2024calibration,ying2025icc}. RT accuracy depends on 3D environment model fidelity. At higher frequencies with shorter wavelengths, objects that are electromagnetically insignificant at lower frequencies become significant scatterers and reflectors, requiring higher spatial resolution in environment models for accurate propagation prediction.

Accurate 3D environment models with material labels present different challenges for outdoor versus indoor scenarios. Outdoor RT simulations can leverage publicly available 3D databases such as OpenStreetMap~\cite{blosm}, enabling open source tools like NVIDIA Sionna~\cite{hoydis2023sionna} to demonstrate propagation prediction without manual reconstruction, but notably also without the ability to calibrate the ray tracer with field measurements of the spatio-temporal channel impulse response~\cite{ying2025site}. Indoor environments lack open source databases and require engineers to reconstruct 3D models manually or with semi-automation using software tools such as Blender, SketchUp, or AutoCAD~\cite{rappaport2005BDM}. Engineers measure room dimensions, place furniture models, and assign material properties based on visual inspection or datasheets~\cite{rappaport2005BDM}. Manual reconstruction for a typical factory environment spanning 700 $\text{m}^2$ requires weeks of graduate student time, which creates a bottleneck for conducting large-scale indoor or outdoor-to-indoor RT simulations.

Researchers have developed several automated reconstruction methods to address the manual bottleneck for creating accurate 3D environment models. RT-specific methods include Building Database Manipulator (BDM)~\cite{rappaport2005BDM,seidel1994tvt}, reception surface-based RT~\cite{rappaport2004RTsurface}, RGB-D sensor-aided reconstruction~\cite{Norisato2023access}, point cloud-based approaches~\cite{Wataru2022pimrc,Qi2023point}, and LiDAR-based environment reconstruction~\cite{zhang2025impact}. However, existing geometry-based methods require manual material labeling or point cloud semantic segmentation, limiting automation. Wireless RT requires both geometric accuracy and precise material characterization through relative permittivity and conductivity values that govern reflection coefficients and transmission losses.

Unlike geometry-based RT, neural field methods such as
NeRF2~\cite{zhao2023nerf2} and Gaussian splatting~\cite{wen2025wrf} use ML to predict wireless channels directly from sparse measurements, bypassing explicit 3D environment modeling. However, neural field approaches require site-specific measurements and retraining for each new environment, limiting practical deployment since network planning requires predictions before site measurements exist. While future computing advances may reduce training time, the inherent dependence on site-specific measurements prevents these methods from generating predictions for new environments prior to deployment.

This paper presents Holistic Reconstruction with Automated Material Assignment (HoRAMA) for Ray Tracing using NYURay, which addresses the aforementioned limitations by automating RT-compatible 3D model generation from standard RGB video that is easily captured on a smartphone or standard low-cost portable camera. HoRAMA combines dense point cloud reconstruction with VLM-aided material labeling to produce geometrically accurate 3D models with automatically embedded material properties in XML format. The major contributions of this paper include:
\begin{itemize}
\item \textbf{Automated RT-Compatible 3D Reconstruction}. HoRAMA generates point clouds from RGB video using real-time dense Simultaneous Localization and Mapping (SLAM) with 3D Reconstruction priors (MASt3R-SLAM)~\cite{murai2025mast3rslam}, followed by mesh simplification and surface smoothing to create manifold mesh geometry, thereby eliminating manual or semi-automated steps required by existing 3D reconstruction methods for wireless RT.

\item \textbf{VLM-Based Material Classification}. HoRAMA uses PTv3~\cite{wu2024ptv3} to segment the point cloud into object instances (walls, tables, chairs). For each segmented object, Qwen3-VL~\cite{Qwen2.5-VL} analyzes corresponding video frames to classify materials (concrete, wood, metal, glass). Material labels map to ITU-R P.2040 EM properties, with majority voting across multiple video frames for robust classification.

\item \textbf{Dual-Band Validation}. This work validates HoRAMA through end-to-end RT simulation accuracy against field measurements. Validation compares RT-simulated versus measured multipath power at 6.75 GHz and 16.95 GHz using two 3D models: manual baseline (laser rangefinder, Blender, visual material inspection by an expert) versus HoRAMA (automated from video). NYURay simulates both 3D models at identical TX-RX measurement locations. RMSE compares matched MPCs~\cite{ying2025gc} in omnidirectional PDPs with antenna gains removed. HoRAMA achieves 2.28 dB RMSE comparable to manual reconstruction (2.18 dB).

\end{itemize}

The remainder of this paper is organized as follows. Section \ref{sec:problem} defines requirements for 3D modeling including geometric accuracy and material properties. Section \ref{sec:method} compares manual reconstruction using Blender with HoRAMA reconstruction from video. Section \ref{sec:validation} validates both reconstruction approaches by comparing NYURay RT simulations against field measurements at 6.75 GHz and 16.95 GHz. Section \ref{sec:conclusion} concludes the paper.

\section{Challenges in Automating 3D Environment Model Creation for Ray Tracing}
\label{sec:problem}

\begin{figure}[!t]
    \centering
    \includegraphics[width=0.8\linewidth]{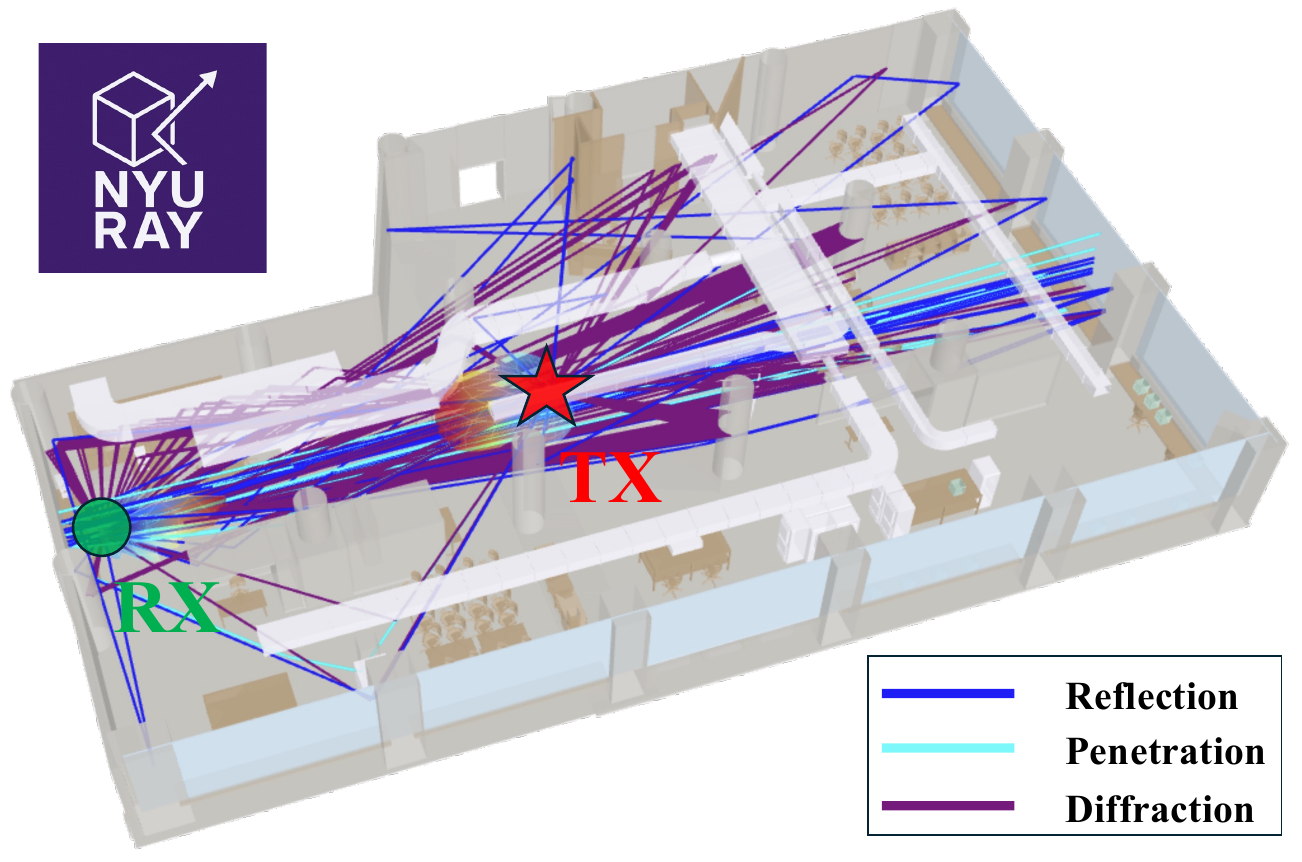}
    \caption{Ray tracing simulation using NYURay in the NYU MakerSpace factory using a material-labeled 3D environment model, where each surface carries a material label (e.g., concrete, glass, metal, wood) mapped to frequency-dependent EM properties via ITU-R P.2040~\cite{ITU-P2040-3}. Traced ray paths from transmitter (red star) to receiver (green circle) demonstrate reflection, penetration, and diffraction propagation mechanisms.}
    \label{fig:rt_example}
    \vspace{-15pt}
\end{figure}

RT-based propagation prediction requires accurate 3D geometry and precise material electromagnetic (EM) characterization. Geometric inaccuracies cause incorrect delays and missing multipath components (MPCs), while material misclassification introduces 3-10 dB power errors per MPC interaction. Fig.~\ref{fig:rt_example} illustrates ray propagation in an indoor factory environment, where numerous MPC interactions can lead to substantial propagation prediction errors.

\subsection{Geometric Accuracy Requirements}

RT algorithms based on the shooting-and-bouncing-ray (SBR) method~\cite{schaubach1992microcellular,seidel1994tvt} require three critical features in the 3D environment model for accurate propagation prediction: (1) spatial positioning accuracy for correct ray-surface intersections, (2) smooth surface representation for accurate reflection angle computation, and (3) watertight manifold mesh topology for unambiguous inside/outside determination at each surface. 
When an incident ray with direction $\mathbf{d}$ strikes a surface, the reflected ray direction $\mathbf{r}$ follows
$\mathbf{r} = \mathbf{d} - 2(\mathbf{d} \cdot \mathbf{n})\cdot\mathbf{n}$, where $\mathbf{d}$, $\mathbf{r}$, and $\mathbf{n}$ denote unit vectors for the incident direction, reflected direction, and surface normal, respectively.

Surface waviness produces diffuse scattering rather than specular reflection, weakening deterministic path prediction~\cite{ying2025site,degli2007measurement}. A watertight manifold mesh is one in which every edge is shared by exactly two triangles and all face normals are consistently oriented, so that the surface unambiguously separates interior from exterior space~\cite{kazhdan2013screened}. RT engines rely on this property to determine whether a ray has entered or exited a solid object at each intersection, which governs whether reflection or penetration coefficients are applied. Non-manifold artifacts, such as holes, self-intersections, or inconsistent normals, cause rays to terminate incorrectly or traverse undefined interiors. 

In this work, 3D environment models are represented as triangulated meshes stored in XML format compatible with the Mitsuba rendering framework~\cite{mitsuba_blender_addon}, which NYURay uses as its input format~\cite{ying2025site,ying2025nyu}. Each triangle face carries a material label that maps to frequency-dependent EM properties ($\varepsilon_r$, $\sigma$) from the ITU-R P.2040 database~\cite{ITU-P2040-3}, enabling the RT engine to retrieve the correct Fresnel reflection and transmission coefficients at every ray-surface intersection.

\subsection{Material Electromagnetic Characterization}

Material EM properties govern wave interactions at surfaces. Complex relative permittivity
\begin{equation}
\eta(\omega) = \varepsilon_r(\omega) - j\frac{\sigma(\omega)}{\varepsilon_0\omega}
\end{equation}
characterizes material response at angular frequency $\omega = 2\pi f$, where $\varepsilon_r$ is relative permittivity, $\sigma$ is conductivity (S/m), and $\varepsilon_0 = 8.854 \times 10^{-12}$ F/m~\cite{rappaport2024wirelessbook}.

Fresnel reflection coefficients compute reflected power at material interfaces. For waves incident from air onto material with $\eta_{mat} = \eta(\omega)$, the reflection coefficients are
\begin{equation}
\begin{split}
r_\perp &= \frac{\cos\theta_i - \sqrt{\eta_{mat} - \sin^2\theta_i}}{\cos\theta_i + \sqrt{\eta_{mat} - \sin^2\theta_i}}, \\
r_\parallel &= \frac{\eta_{mat}\cos\theta_i - \sqrt{\eta_{mat} - \sin^2\theta_i}}{\eta_{mat}\cos\theta_i + \sqrt{\eta_{mat} - \sin^2\theta_i}}
\end{split}
\end{equation}
where $\theta_i$ denotes the incident angle measured from the surface normal. Reflected power fraction equals $|r|^2$, while transmission coefficients $t = 1 + r$ determine penetration power~\cite{rappaport2024wirelessbook}.

Material misclassification directly impacts power prediction. For example, misclassifying concrete ($\varepsilon_r = 5.24$, $\sigma = 0.237$ S/m) as wood ($\varepsilon_r = 1.99$, $\sigma = 0.036$ S/m) at 6.75 GHz produces 3-5 dB reflection error and 6-10 dB penetration error. Power prediction errors accumulate across multiple interactions~\cite{ITU-P2040-3}.

\subsection{Validation Metric}

In this work, reconstruction quality is quantified by power delay profile (PDP) RMSE, when comparing matched MPCs in RT-predicted and measured omnidirectional PDPs, computed over an ensemble of TX--RX measurement location pairs. For $N$ matched multipath component pairs, i.e., RT-simulated MPCs paired one-to-one with measured MPCs that share the same physical propagation path, identified by minimizing a weighted delay-power distance~\cite{ying2025gc} (see Section~\ref{sec:validation}), RSS values are converted from dBm to linear scale (milliwatts). RMSE is computed in linear scale as
\begin{equation}
    \label{eq:RMSE}
\text{RMSE}_{linear} = \sqrt{\frac{1}{N}\sum_{i=1}^N (P_{RT,i} - P_{meas,i})^2}
\end{equation}
where $P_{RT,i}$ and $P_{meas,i}$ denote the RT-predicted and measured received power of the $i$-th matched MPC, and $N$ denotes the total number of matched MPC pairs aggregated over all TX--RX measurement locations. Linear RMSE converts to dB as
\begin{equation}
\text{RMSE}_{dB} = 10 \log_{10}(\text{RMSE}_{linear}).
\end{equation}
We validate HoRAMA by comparing reconstruction methods with measurements at both 6.75 GHz and 16.95 GHz in Section~\ref{sec:validation}.

\section{3D Environment Reconstruction Methods}
\label{sec:method}
This section presents and compares two 3D environment model reconstruction approaches: manual baseline reconstruction by a human expert and HoRAMA automated reconstruction. Both methods generate RT-compatible 3D models with geometric structure and material labels for the NYU MakerSpace factory environment, which has been extensively measured and reported in prior FR3 and mmWave channel measurement studies~\cite{Shakya2024ojcoms,ying2025icc,kanhere2024calibration}. Both produce the same output: a triangulated mesh in Mitsuba XML format with per-face material labels that map to frequency-dependent EM properties ($\varepsilon_r$, $\sigma$) from the ITU-R P.2040 database~\cite{ITU-P2040-3}.

\subsection{Manual 3D Reconstruction}

Manual 3D reconstruction serves as the baseline, using Blender CAD modeling guided by systematic field measurements and floor plans, with the resulting geometry exported as RT-compatible surface meshes with material labels for use in NYURay. The Bosch DLR165 Laser Rangefinder (1.5 mm accuracy) captures room dimensions, wall thickness, and ceiling height. The NYU MakerSpace 3D modeling focuses on electromagnetically significant objects including 3D printers, mechanical machines (CNC, laser cutters), tables, chairs, windows, monitors, and cabinets. Each element is manually created in Blender using primitive geometric shapes (cubes, cylinders, planes) modified through extrusion, scaling, and boolean operations to match measured dimensions.

Material labels are assigned to each object based on on-site visual inspection and available manufacturer datasheets, reflecting the actual construction and equipment materials of the NYU MakerSpace factory. This process was conducted by an experienced graduate researcher with prior wireless propagation measurement and 3D modeling experience. The environment consists primarily of concrete walls, glass windows, metal machine frames and fixtures, and wooden tables and chairs. Material EM properties are retrieved from ITU-R P.2040 database~\cite{ITU-P2040-3}. The manual 3D reconstruction of the 700~$\text{m}^2$ factory, including on-site geometric measurements, CAD modeling in Blender, and manual material labeling for RT, required approximately two months by an experienced operator, achieving centimeter-level geometric accuracy for major structural elements and equipment placement. Fig.~\ref{fig:rt_example} illustrates RT simulation using the manual 3D environment model, which serves as the baseline for validating HoRAMA's automated output in the identical Mitsuba XML format.

\subsection{HoRAMA Automated 3D Reconstruction}

\begin{figure}[!t]
    \centering
    \includegraphics[width=0.78\linewidth]{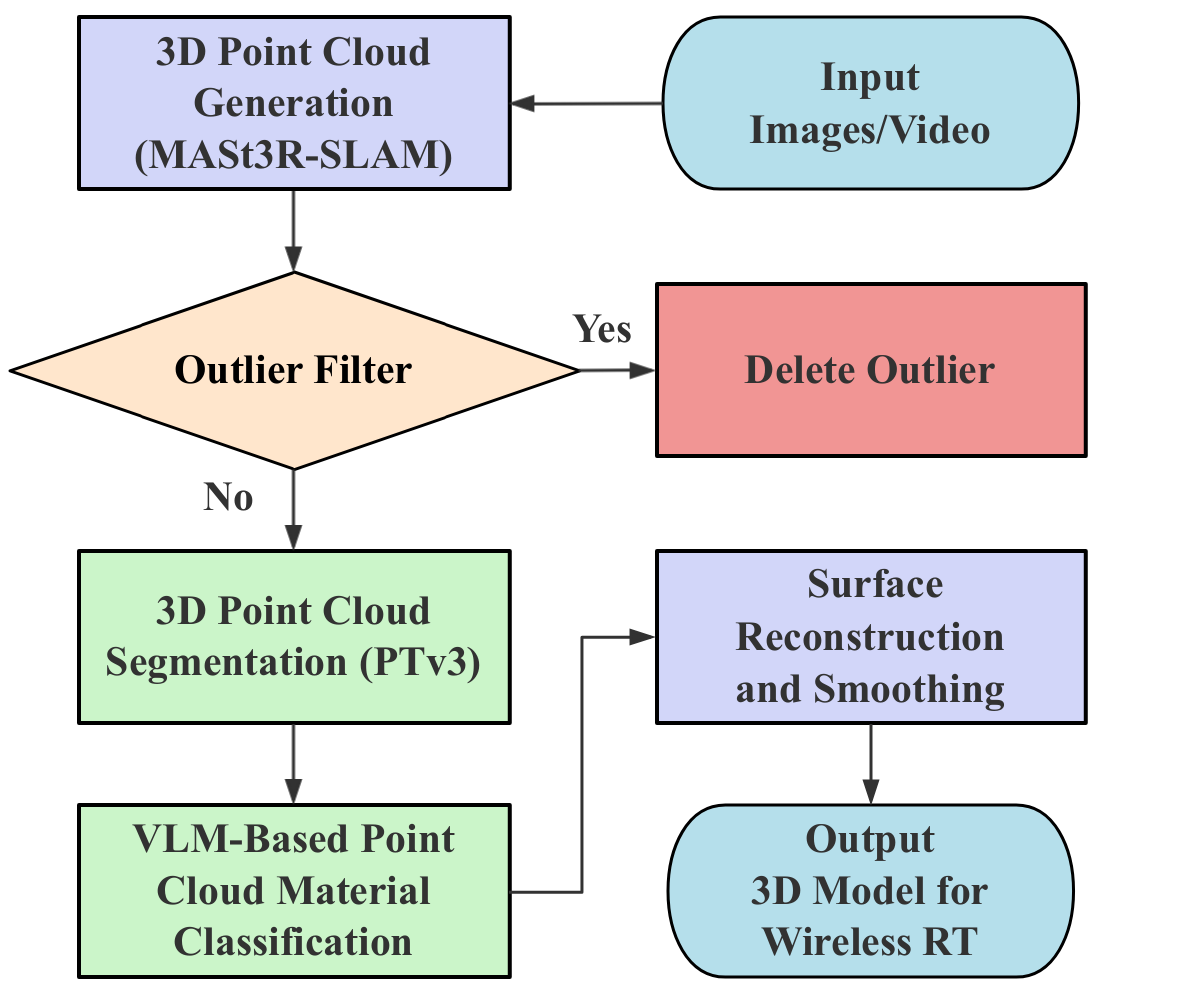}
    \caption{HoRAMA workflow for automated 3D reconstruction and material classification from RGB video. The pipeline processes video through dense point cloud generation (MASt3R-SLAM), semantic segmentation (PTv3), VLM material classification (Qwen3-VL), and surface reconstruction.}
    \label{fig:flowchart}
    \vspace{-15pt}
\end{figure}

\begin{figure*}[!t]
    \centering
    \begin{subfigure}[b]{0.25\textwidth}
        \centering
        \includegraphics[height=0.13\textheight,keepaspectratio]{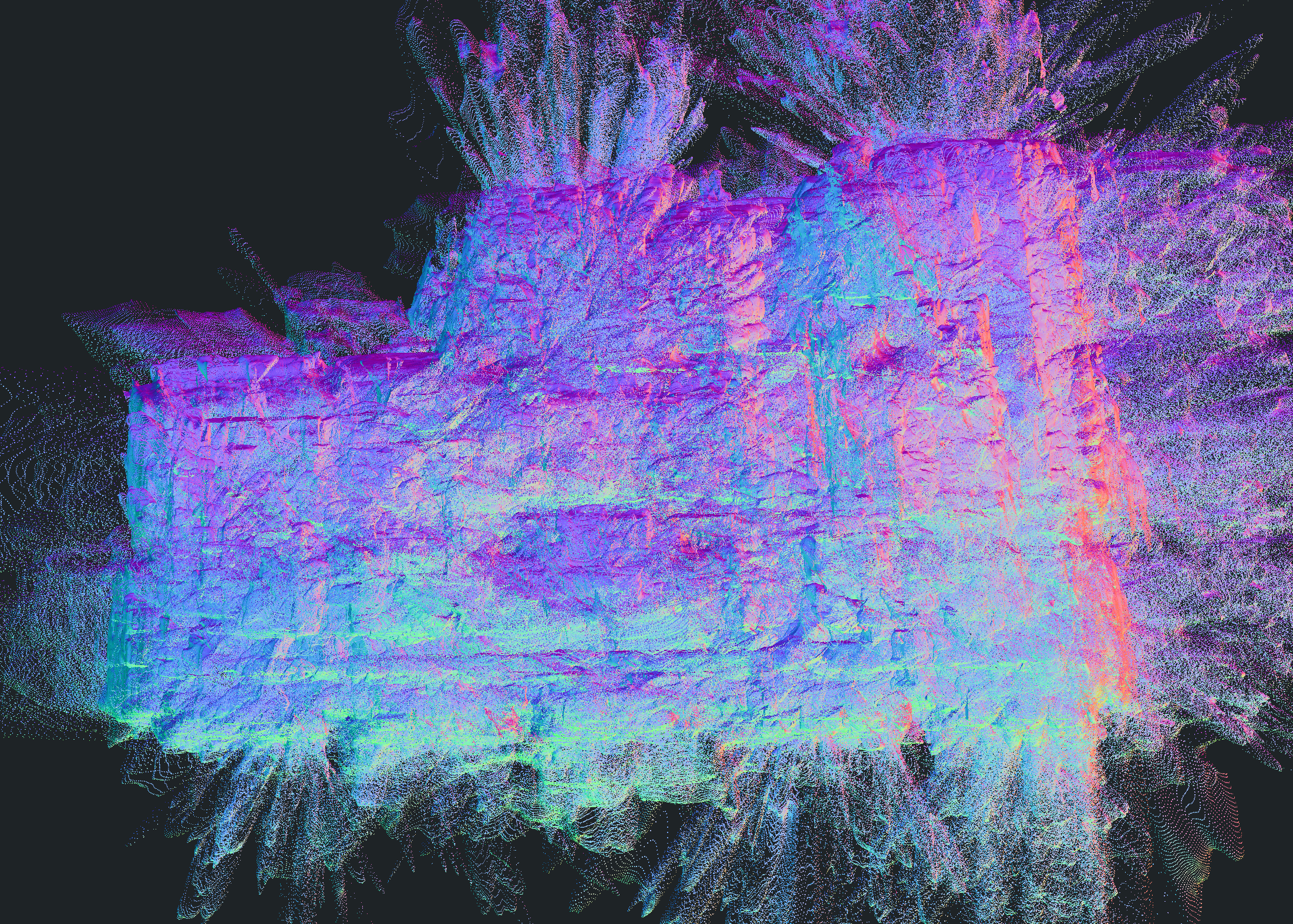}
        \vspace{-5pt}
        \caption{}
    \end{subfigure}\hfill%
    \begin{subfigure}[b]{0.25\textwidth}
        \centering
        \includegraphics[height=0.13\textheight,keepaspectratio]{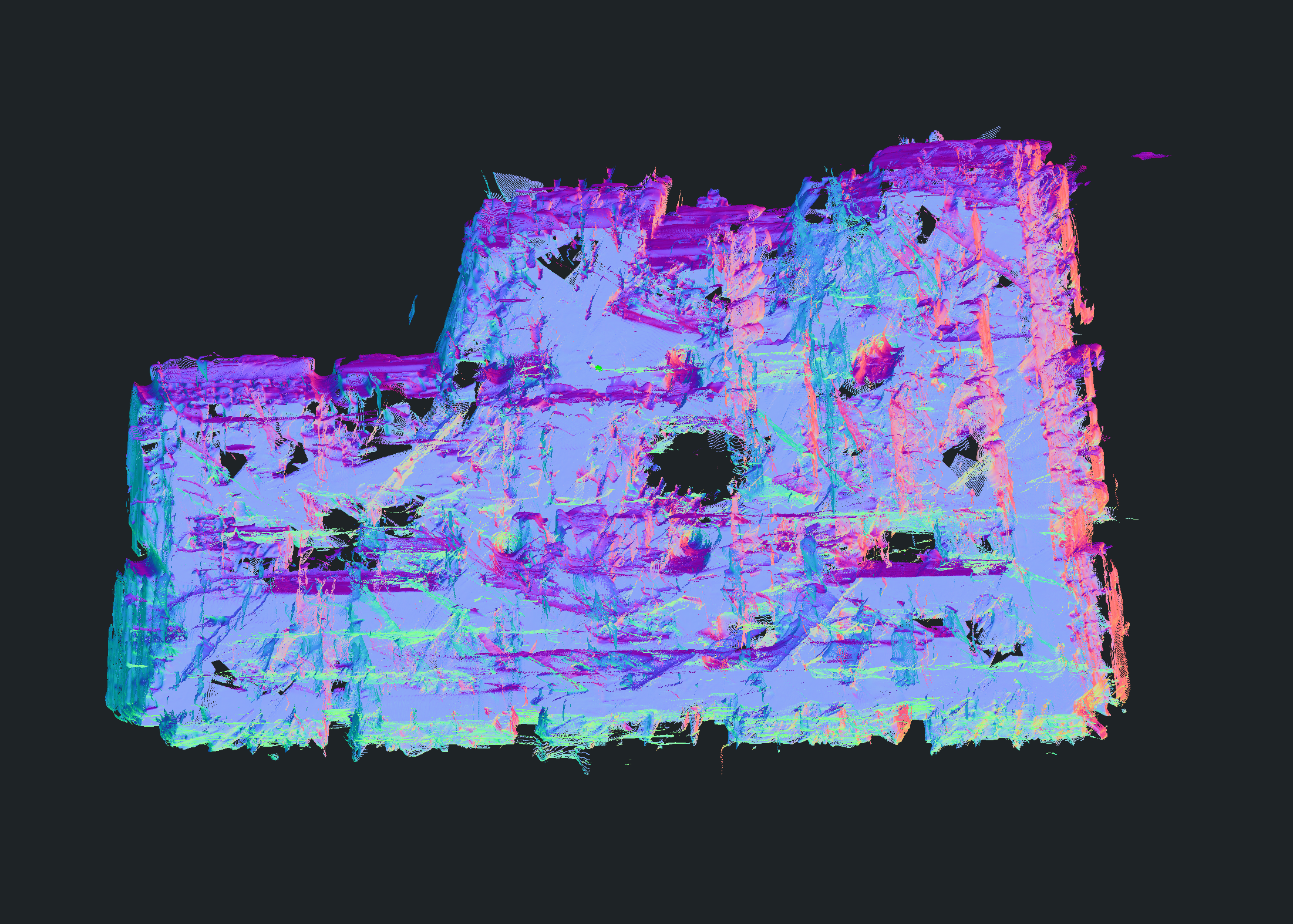}
        \vspace{-5pt}
        \caption{}
    \end{subfigure}\hfill%
    \begin{subfigure}[b]{0.25\textwidth}
        \centering
        \includegraphics[height=0.13\textheight,keepaspectratio]{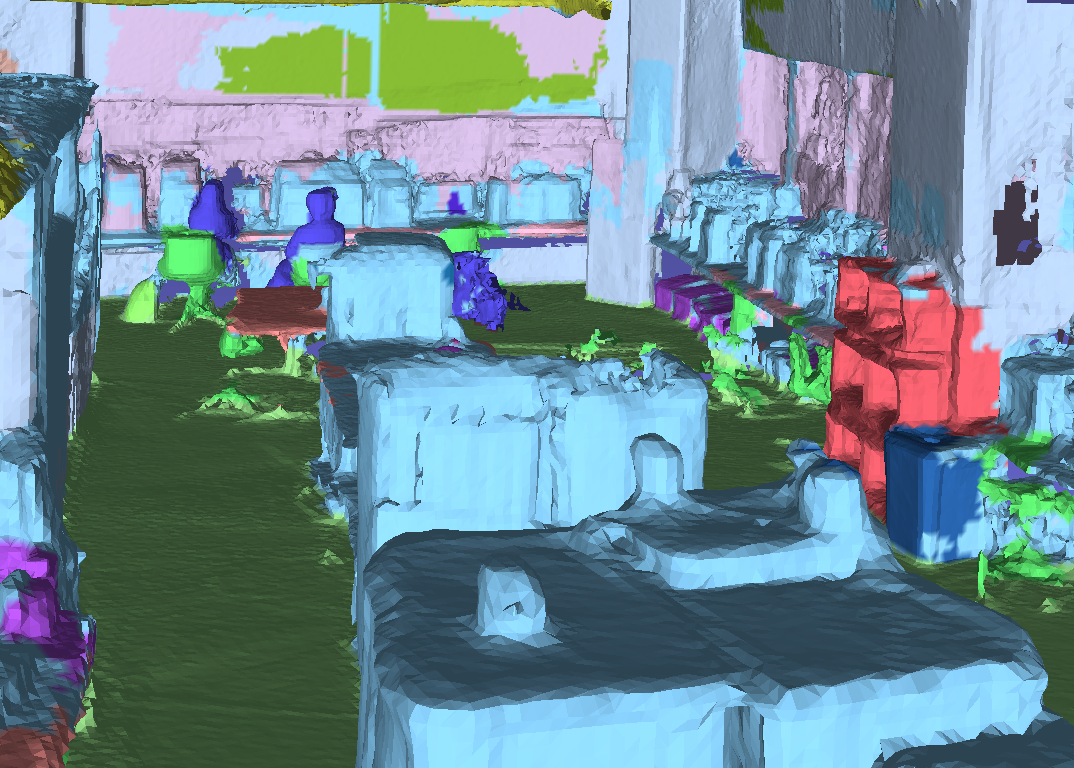}
        \vspace{-5pt}
        \caption{}
    \end{subfigure}\hfill%
    \begin{subfigure}[b]{0.25\textwidth}
        \centering
        \includegraphics[height=0.13\textheight,keepaspectratio]{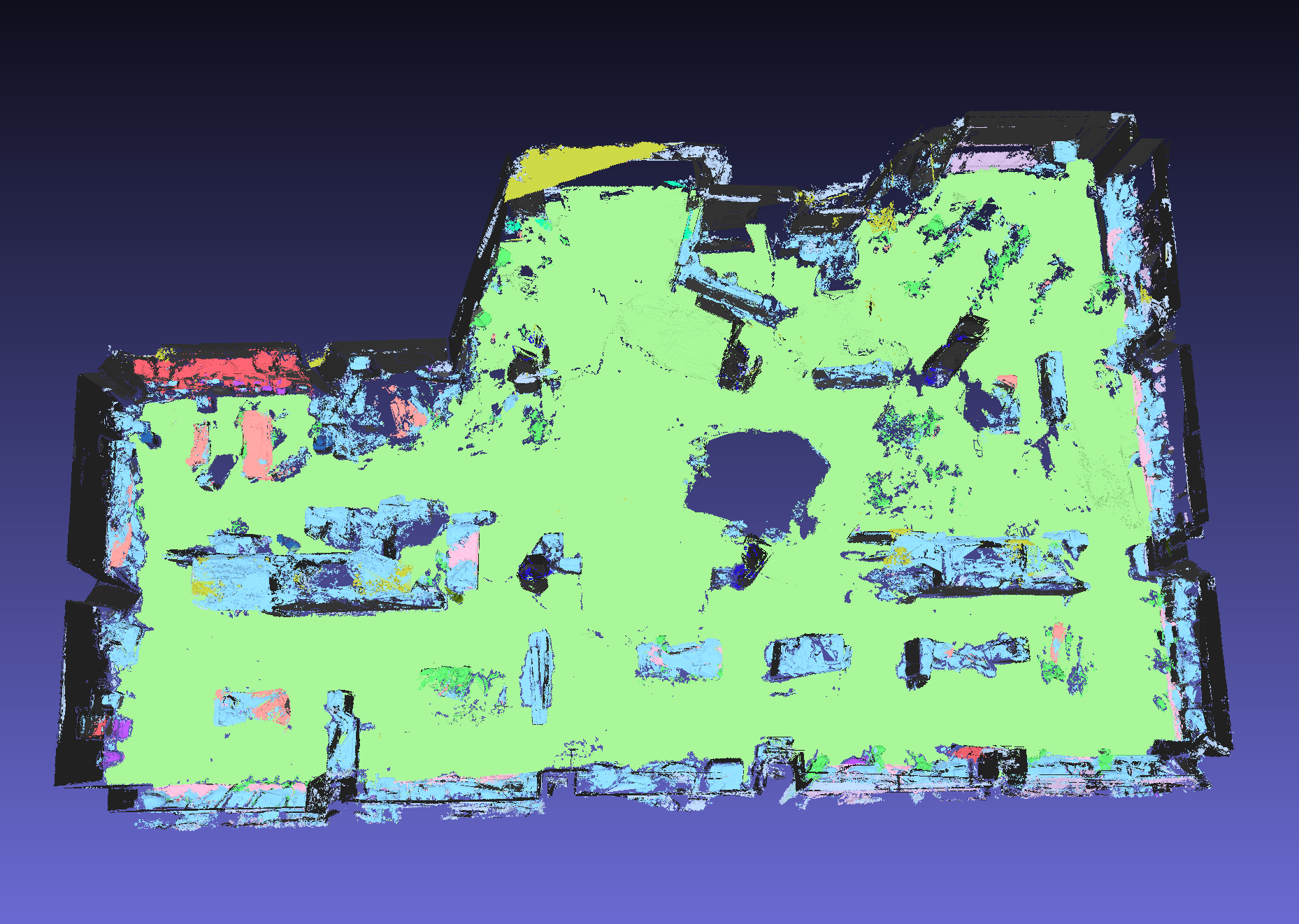}
        \vspace{-5pt}
        \caption{}
    \end{subfigure}
    \vspace{-18pt}
    \caption{HoRAMA automated 3D reconstruction from raw point cloud to RT-ready model: (a) Raw dense point cloud from MASt3R-SLAM with color-coded surface normals with noise and outliers. (b) Cleaned point cloud with color-coded surface normals after outlier filtering. (c) Partial view of PTv3 segmented indoor factory environment before smoothing and surface reconstruction. (d) Final 3D environment map with smooth planar surfaces and VLM-classified material labels from ITU-R P.2040 categories (concrete, wood, metal, glass, plywood).}
    \label{fig:reconstruction_stages}
    \vspace{-14pt}
\end{figure*}

HoRAMA automates 3D model generation from RGB video footage, replacing the laser rangefinder measurements, Blender CAD modeling, and on-site material inspection described in Section~III-A with a vision-based pipeline that produces the same RT-ready Mitsuba XML output. Fig.~\ref{fig:flowchart} illustrates the complete HoRAMA workflow. Fig.~\ref{fig:reconstruction_stages} shows progressive refinement stages from raw point clouds with color-coded surface normals through outlier filtering, intermediate reconstruction, to the final RT-ready 3D environment map with smooth surfaces and material labels.

\textbf{Input Stage.} Video acquisition uses a commodity camera (iPhone 16 Pro) capturing RGB video at 1920$\times$1440 resolution at 60 fps. An operator walks through the environment at approximately 1~m/s, panning the camera to capture all surfaces from multiple viewing angles. Typical video duration for the 700 $\text{m}^2$ factory is 10 minutes ($\sim$1~GB, H.265), capturing 36000 frames.

\textbf{3D Point Cloud Generation.} MASt3R-SLAM~\cite{murai2025mast3rslam} is a real-time monocular dense SLAM system that takes the RGB video as input and produces globally consistent camera poses and dense 3D pointmaps $\mathbf{X} \in \mathbb{R}^{H\times W\times3}$ with per-pixel confidence scores $\mathbf{C} \in \mathbb{R}^{H\times W}$ for each frame, where $H$ and $W$ denote image height and width. MASt3R-SLAM is chosen because it operates on uncalibrated monocular video without requiring depth sensors or known camera intrinsics, matching the commodity smartphone capture described above. When the same 3D point is observed from multiple frames, MASt3R-SLAM fuses these observations into a single estimate through confidence-weighted averaging, and applies loop closure with global optimization to reduce accumulated drift over the walk-around trajectory~\cite{murai2025mast3rslam}.

\textbf{Outlier Filtering.} The raw pointmaps $\mathbf{X}$ and confidence scores $\mathbf{C}$ from MASt3R-SLAM contain outliers from transient objects, specular reflections, and matching errors. Multi-stage filtering removes unreliable points while preserving valid geometry, producing cleaned point cloud $\mathcal{P} = \{\mathbf{p}_i\}_{i=1}^{M}$ where $M$ denotes the number of valid 3D points. Robust Huber weighting downweights outlier residuals through
\begin{equation}
w(r) = \min \{1, k/|r| \}
\end{equation}
where $r = \|\mathbf{p}_{\text{predicted}} - \mathbf{p}_{\text{observed}}\|$ denotes the reprojection error between predicted and observed 3D point positions, and $k = 1.345$ denotes the Huber threshold. Geometric distance filtering rejects 3D points with reprojection distance exceeding 0.1 meters. This paper suggests an automated pipeline that fuses multiple estimated 3D coordinates (observations) of the same 3D point position through confidence-weighted averaging
\begin{equation}
\bar{\mathbf{p}} = \frac{\sum_{k=1}^N w_k \mathbf{p}_k}{\sum_{k=1}^N w_k}, \quad w_k = C_k \cdot Q_k
\end{equation}
where $\mathbf{p}_k$ denotes the $k$-th observation, $N$ denotes observation count, $C_k$ denotes confidence score, and $Q_k$ denotes quality score. Size filtering removes outlier points and noise by eliminating small clusters that contain fewer than 100 points, retaining 85-90\% of original points. Fig.~\ref{fig:reconstruction_stages}(a,b) visualizes raw and cleaned point clouds with color-coded surface normals.

\textbf{Point Cloud Segmentation.} The cleaned point cloud $\mathcal{P} = \{\mathbf{p}_i\}_{i=1}^{M}$ from the previous stage contains $M$ 3D surface points, where each point $\mathbf{p}_i \in \mathbb{R}^3$ represents an $(x,y,z)$ position on a physical surface (e.g., a wall, table, or machine) in the environment. PTv3~\cite{wu2024ptv3}, a deep learning model trained on the S3DIS indoor dataset (6020~$\text{m}^2$, 12 semantic classes), classifies each point into a semantic category $\ell_i \in \{\text{wall, floor, ceiling, table, chair, \ldots}\}$, assigning a material-relevant label to every surface point in the cloud. Euclidean clustering then groups spatially adjacent points sharing the same label into distinct object instances $O = \{o_1, o_2, \ldots, o_K\}$; for example, two separate tables in the factory become two distinct instances, each receiving its own material label in the next stage.

\textbf{VLM-Based Material Classification.}
Each object instance $o_k$ (e.g., a specific table or wall segment) from the segmentation stage is a cluster of 3D points that must be assigned a material label for RT. Since MASt3R-SLAM maintains the correspondence between each 3D point and its source RGB pixel in the original video, HoRAMA identifies which video frames contain object $o_k$ and crops the corresponding RGB image region from each frame. Qwen3-VL~\cite{Qwen2.5-VL}, a vision-language model, examines each cropped image region and selects a material from the ITU-R P.2040 categories (concrete, wood, metal, glass, plywood). Because the same object appears in multiple video frames from different viewing angles, HoRAMA collects material predictions across all frames where $o_k$ is visible. Let $F(o_k)$ denote the set of frames observing object $o_k$. Final material assignment employs majority voting across all frames in which object $o_k$ appears
\begin{equation}
m^*(o_k) = \underset{m}{\arg\max} \sum_{\text{frame} \in F(o_k)} \mathbf{1}[m_{\text{frame}} = m]
\end{equation}
where $m^*(o_k)$ denotes the final material label for object $o_k$, and $\mathbf{1}[\cdot]$ denotes the indicator function that equals 1 when the condition is true and 0 otherwise. Majority voting improves robustness against occlusions and lighting variations. Visual classification determines surface material but cannot infer thickness or internal structure. Material labels map to frequency-dependent complex permittivity and conductivity from ITU-R P.2040, with nominal parameters that field calibration can refine~\cite{kanhere2024calibration}.

\textbf{Surface Reconstruction and Smoothing.} The material-labeled object instances from the previous stages are now converted from point clouds into triangle meshes for RT. For planar objects (walls, floors, ceilings), we apply PLANA3R~\cite{plana3r2025} to extract planar surface primitives from the points belonging to each object instance. Indoor environments are dominated by large flat surfaces, and PLANA3R represents these as compact planar primitives rather than dense point clouds~\cite{plana3r2025}. Adjacent primitives with projection distance $\leq 0.1$~m and normal angle difference $\leq 10^\circ$ (adapted from PLANA3R~\cite{plana3r2025} with a tighter normal threshold to enforce surface flatness for accurate specular reflection in RT) are merged, then fitted to plane equation $\mathbf{n}^\top \mathbf{p} + d = 0$ using area-weighted least squares. For non-planar objects (furniture, machines), Poisson surface reconstruction~\cite{kazhdan2013screened} generates watertight manifold meshes. Finally, planar mesh vertices project onto fitted plane references, enforcing geometric flatness required for accurate specular reflection in RT (Fig.~\ref{fig:reconstruction_stages}(c,d)). 

\textbf{Output Stage.}   Material labels attach to mesh faces, enabling RT engines to retrieve EM properties.   HoRAMA exports the 3D model in XML format compatible with standard RT engines using Mitsuba. With 15 fps video downsampling, the 700 $\text{m}^2$ indoor factory reconstruction completes within 16 hours on dual NVIDIA RTX 4090 GPUs (24GB VRAM each) with AMD Ryzen Threadripper PRO 7975WX 32-core CPU, compared to approximately two months required for manual 3D reconstruction by an experienced operator.

\section{Validation}
\label{sec:validation}

Comprehensive validation in the InF scenario quantifies the accuracy of each reconstruction approach through dual-band channel measurements and RT simulations in a representative indoor factory environment. The validation methodology establishes ground truth measured channels at identical TX-RX locations where RT simulations are performed, enabling direct RMSE comparison across different reconstruction methods.

\subsection{Validation Setup}

\subsubsection{Measurement Setup}
\begin{figure}[!t]
    \centering
    \includegraphics[width=0.99\linewidth]{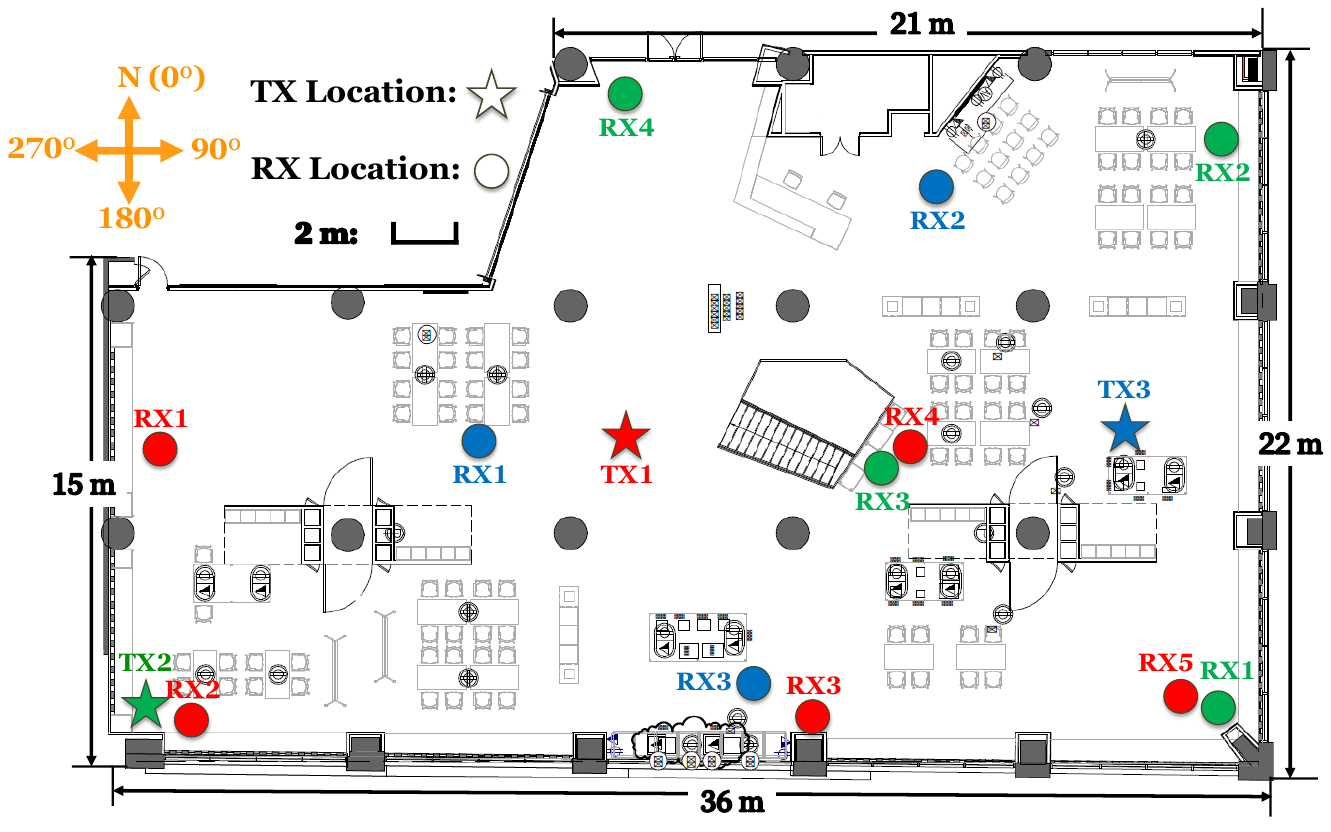}
    \caption{NYU MakerSpace Factory floor plan showing 12 TX-RX location pairs for channel measurements at 6.75 GHz and 16.95 GHz. Stars represent TX locations, and matching colored circles represent corresponding RX locations for each TX.}
    \label{fig:factory_map}
    \vspace{-5pt}
\end{figure}

The validation environment comprises the NYU MakerSpace Factory located on the 1st floor of 6 MetroTech, Brooklyn, NY, spanning \SI{36}{\meter} $\times$ \SI{22}{\meter} $\times$ \SI{5}{\meter} (W$\times$L$\times$H), covering approximately 700 $\text{m}^2$. The MakerSpace contains diverse machinery with exterior walls featuring large glass panels framed with metal and concrete pillars distributed throughout, creating typical indoor factory (InF) scenarios~\cite{ying2025icc}. Fig.~\ref{fig:factory_map} illustrates the factory layout with 5 LOS and 7 NLOS TX-RX location pairs (T-R separation: 9-38 m). The channel sounder employs wideband sliding-correlation operating at \SI{6.75}{\giga\hertz} and \SI{16.95}{\giga\hertz} with \SI{1}{\giga\hertz} bandwidth, achieving \SI{1}{\nano\second} multipath resolution~\cite{Shakya2024ojcoms,ying2025icc}. Pyramidal horn antennas providing 15 dBi gain at \SI{6.75}{\giga\hertz} and 20 dBi gain at \SI{16.95}{\giga\hertz} were used at TX (\SI{2.4}{\meter}) and RX (\SI{1.5}{\meter}) heights. Both antennas performed azimuth and elevation scans in half-power beamwidth (HPBW) increments, synthesizing omni-PDPs by summing power across all directional PDPs. 

\subsection{PDP Comparison Methodology}
\begin{table}[!t]
\centering
\caption{Ray-Tracing Simulation Parameters}
\vspace{-3pt}
\label{tab:simulation_parameters}
\renewcommand{\arraystretch}{1}
\begin{tabular}{p{3cm}p{4cm}}
\toprule
\textbf{Parameter} & \textbf{Value} \\
\midrule
Frequencies & \SI{6.75}{\giga\hertz}, \SI{16.95}{\giga\hertz} \\
TX Power & 0 dBm \\
TX / RX Height & \SI{2.4}{\meter}\,/\,\SI{1.5}{\meter} \\
TX / RX Antenna Pattern & Isotropic (0 dBi) \\
TX / RX Polarization & Vertical \\
Number of Rays & $10^6$ (Fibonacci lattice) \\
Max Num of Reflection & 5 \\
Propagation Mechanisms & Reflection, Penetration, Diffraction \\
Materials Characteristics& ITU-R P.2040~\cite{ITU-P2040-3} \\
\bottomrule
\end{tabular}
\vspace{-10pt}
\end{table}

\subsubsection{RT Simulation using NYURay}

RT simulations employ NYURay, an in-house hybrid shooting-bouncing-ray and image-based RT engine~\cite{ying2025site}. At each ray-surface intersection, NYURay retrieves surface normal and material properties ($\varepsilon_r$, $\sigma$) to compute Fresnel coefficients via Equations (2) and (3) in Section~\ref{sec:problem}. Reflected rays follow directions determined by surface normals with power $P_{reflected} = |r|^2 P_{incident}$.

Table~\ref{tab:simulation_parameters} summarizes RT simulation parameters. One million rays launch from each TX location using Fibonacci lattice sampling over the unit sphere in three-dimensional space. Of the $10^6$ launched rays, only those whose propagation paths geometrically intersect the point RX location yield valid MPCs; for each TX-RX pair in this indoor factory scenario, approximately one hundred valid MPCs reach the RX, with tens from reflections alone and additional MPCs contributed by diffraction and penetration. Material characteristics are retrieved from ITU-R P.2040 database~\cite{ITU-P2040-3}. Each TX-RX pair requires approximately 5 seconds to compute. Simulations are executed independently for both environment models (manual and HoRAMA) at all 12 TX-RX locations and at both frequency bands. Each run produces an omnidirectional PDP containing multipath delay and power.

\begin{table}[!t]
    \centering
    \caption{RMSE (dB) between RT-simulated and measured matched MPC powers~\cite{ying2025gc} in omnidirectional PDPs for manual 3D reconstruction and HoRAMA.}
    \vspace{-3pt}
    \renewcommand{\arraystretch}{1.28}
    \label{tab:rmse_comparison}
    \footnotesize
    \setlength{\tabcolsep}{3.5pt}
    \begin{tabular}{llcccc}
    \toprule
    \textbf{Freq.} & \textbf{Scenario} & \multicolumn{2}{c}{\textbf{Manual 3D (dB)}} &
    \multicolumn{2}{c}{\textbf{HoRAMA (dB)}} \\
    \textbf{(GHz)} & & \multicolumn{2}{c}{Mean \textpm\ Std. \ \ Med.} &
    \multicolumn{2}{c}{Mean \textpm\ Std. \ \ Med.} \\
    \midrule
    \multirow{3}{*}{6.75} & LOS & 0.40 \textpm\ 0.46 & 0.24 & 0.40 \textpm\ 0.38 & 0.24 \\
     & NLOS & 3.65 \textpm\ 1.71 & 4.05 & 3.54 \textpm\ 1.71 & 3.40 \\
     & \cellcolor{gray!20}Overall & \cellcolor{gray!20}2.17 \textpm\ 2.10 &
    \cellcolor{gray!20}1.41 & \cellcolor{gray!20}1.54 \textpm\ 1.86 & \cellcolor{gray!20}0.54
    \\
    \midrule
    \multirow{3}{*}{16.95} & LOS & 0.59 \textpm\ 0.39 & 0.40 & 1.89 \textpm\ 1.37 & 1.71 \\
     & NLOS & 3.27 \textpm\ 1.41 & 3.09 & 4.47 \textpm\ 4.30 & 2.20 \\
     & \cellcolor{gray!20}Overall & \cellcolor{gray!20}2.20 \textpm\ 1.75 &
    \cellcolor{gray!20}2.09 & \cellcolor{gray!20}2.96 \textpm\ 3.09 & \cellcolor{gray!20}2.14
    \\
    \midrule
    \multirow{3}{*}{Combined} & LOS & 0.48 \textpm\ 0.41 & 0.38 & 1.14 \textpm\ 1.24 & 0.43 \\
     & NLOS & 3.46 \textpm\ 1.51 & 3.29 & 4.06 \textpm\ 3.25 & 2.71 \\
     & \cellcolor{gray!20}\textbf{Overall} & \cellcolor{gray!20}\textbf{2.18 \textpm\ 1.90} &
    \cellcolor{gray!20}1.51 & \cellcolor{gray!20}\textbf{2.28 \textpm\ 2.62} & \cellcolor{gray!20}1.71
    \\
    \bottomrule
    \end{tabular}
     \vspace*{-10pt}
    \end{table}

Validation quantifies reconstruction accuracy by comparing matched MPCs in RT-simulated omnidirectional PDPs against measured omnidirectional PDPs at identical TX-RX locations. RT simulations produce omnidirectional PDPs using isotropic (0 dBi) antenna patterns at both TX and RX, computing received power from all ray propagation paths. Field measurements synthesize omnidirectional PDPs by summing received power across azimuth and elevation antenna scans, with directional antenna gains (15 dBi at 6.75 GHz, 20 dBi at 16.95 GHz) mathematically removed to produce equivalent isotropic reception.

Both RT-simulated and measured PDPs contain multipath components (MPCs), each characterized by propagation delay $\tau$ and received power $P$. Each MPC represents a distinct propagation path through reflection, diffraction, or penetration, and is uniquely defined by its angle of departure (AoD), angle of arrival (AoA), propagation delay, and interaction sequence. MPC extraction applies a 25 dB dynamic range threshold below peak received power to both RT and measurement PDPs, filtering weak components dominated by noise~\cite{ying2025icc}. For each TX-RX pair, NYURay uses SBR to identify candidate multipath sequences and the image method to validate exact geometric paths to the point RX location, with no reception sphere~\cite{ying2025site,ying2025nyu}. Each valid path yields a complex amplitude $a_i$ (with magnitude and phase) and delay $\tau_i$; the omnidirectional PDP is formed non-coherently as $P(\tau) = \sum_i |a_i|^2 \delta(t - \tau_i)$, where each MPC's received power is $|a_i|^2$ and no phasor summation is performed across distinct paths. RT-simulated and measured MPCs are then paired using the weighted delay-power matching method in~\cite{ying2025gc}. A \emph{matched MPC} is an RT-simulated MPC that has been paired with a measured MPC by minimizing a weighted delay-power distance, such that both represent the same physical propagation path. Measurements typically capture more MPCs than RT, since real-world environments contain finer geometric details and richer scattering and diffraction interactions than the simplified 3D model can fully resolve. RMSE quantifies prediction accuracy by comparing received power values of matched RT-simulated versus measured MPCs~\cite{ying2025gc}, as defined in Equation~\eqref{eq:RMSE}. 

\subsection{Comparative Results}

The validation results show that HoRAMA achieves comparable accuracy in omnidirectional power–delay profile (PDP) prediction to the manual 3D reconstruction baseline, as quantified by the RMSE (in dB) between ray-traced and measured matched MPC powers in omnidirectional PDPs~\cite{ying2025gc} (Table~\ref{tab:rmse_comparison}, Eq.~\eqref{eq:RMSE}). Manual 3D model reconstruction achieves 2.18 dB combined overall RMSE across both frequencies. HoRAMA achieves 2.28 dB combined overall RMSE, only 0.10 dB higher than manual reconstruction, but vastly reducing the implementation time from 2 months to 16 hours. LOS scenarios exhibit consistently lower RMSE for both methods compared to NLOS scenarios, reflecting increased multipath complexity in obstructed propagation. Manual 3D model reconstruction achieves 0.48 dB RMSE in LOS scenarios versus 3.46 dB in NLOS scenarios. HoRAMA achieves 1.14 dB RMSE in LOS scenarios versus 4.06 dB in NLOS scenarios.

Frequency-dependent analysis reveals distinct performance characteristics. At 6.75 GHz, HoRAMA achieves 1.54 dB overall RMSE, compared to 2.17 dB for manual reconstruction. At 16.95 GHz, manual reconstruction achieves 2.20 dB while HoRAMA exhibits 2.96 dB. The performance gap at 16.95 GHz (HoRAMA: 2.96 dB vs. manual: 2.20 dB) reflects increased sensitivity to geometric accuracy at shorter wavelengths ($\lambda = 17.7$ mm at 16.95 GHz vs. 44.4 mm at 6.75 GHz). The outlier filtering and surface smoothing stages in HoRAMA remove fine geometric details that manual reconstruction preserves, causing wavelength-scale features such as machine edges and small objects to affect higher-frequency propagation more significantly. The median RMSE values show even closer agreement between methods, with manual reconstruction achieving 1.51 dB and HoRAMA achieving 1.71 dB across both frequencies.

\section{Conclusion}
\label{sec:conclusion}

HoRAMA integrates MASt3R-SLAM dense point cloud generation with PTv3 semantic segmentation and VLM-assisted material classification to produce RT-compatible 3D models from RGB video footage, addressing the scalability bottleneck of manual 3D model reconstruction for wireless RT. Complex RT environment model creation was reduced from 2 months to 16 hours with no loss of accuracy. Dual-band validation demonstrates HoRAMA achieves 2.28 dB RMSE for matched MPCs~\cite{ying2025gc} in omnidirectional PDPs, comparable to manual reconstruction (2.18 dB) while reducing reconstruction time. Future work will validate HoRAMA in diverse environments including outdoor urban scenarios using drone-based video capture, with further work in calibration and accuracy improvement.

\bibliographystyle{IEEEtran}
\bibliography{references}

\end{document}